\documentclass[3p]{elsarticle}

\usepackage{lineno,hyperref}
\usepackage{color}
\usepackage{graphicx}
\usepackage{hyperref}
\usepackage{multirow}
\usepackage{verbatim}
\usepackage{setspace}
\usepackage{fontenc}
\usepackage{amsmath}
\usepackage{caption}
\usepackage{chemfig}
\usepackage{nicefrac}
\modulolinenumbers[5]
\biboptions{sort&compress}
\journal{Materials Science and Engineering: R: Reports}

\begin{document}
	
	\begin{frontmatter}
		
		\title{Polymer Informatics: Current Status and Critical Next Steps}
		
		\author[1]{Lihua Chen}
		\author[2]{Ghanshyam Pilania}
		\author[3]{Rohit Batra}
		\author[1]{Tran Doan Huan}
		\author[1]{Chiho Kim}
		\author[1]{Christopher Kuenneth}
		\author[1]{Rampi Ramprasad\corref{mycorrespondingauthor}}
		\cortext[mycorrespondingauthor]{Corresponding author}
		\ead{rampi.ramprasad@mse.gatech.edu}
		
		\address[1]{School of Materials Science and Engineering, Georgia Institute of Technology, Atlanta, GA 30332, USA}
		\address[2]{Materials Science and Technology Division, Los Alamos National Laboratory, Los Alamos, NM 87545, USA}
		\address[3]{Center for Nanoscale Materials, Argonne National Laboratory, Lemont, Illinois 60439,USA}
		\begin{abstract}
Artificial intelligence (AI) based approaches are beginning to impact several domains of human life, science and technology. Polymer informatics is one such domain where AI and machine learning (ML) tools are being used in the efficient development, design and discovery of polymers. Surrogate models are trained on available polymer data for instant property prediction, allowing screening of promising polymer candidates with specific target property requirements. Questions regarding synthesizability, and potential (retro)synthesis steps to create a target polymer, are being explored using statistical means. Data-driven strategies to tackle unique challenges resulting from the extraordinary chemical and physical diversity of polymers at small and large scales are being explored. Other major hurdles for polymer informatics are the lack of widespread availability of curated and organized data, and approaches to create machine-readable representations that capture not just the structure of complex polymeric situations but also synthesis and processing conditions. Methods to solve inverse problems, wherein polymer recommendations are made using advanced AI algorithms that meet application targets, are being investigated. As various parts of the burgeoning polymer informatics ecosystem mature and become integrated, efficiency improvements, accelerated discoveries and increased productivity can result. Here, we review emergent components of this polymer informatics ecosystem and discuss imminent challenges and opportunities.
		\end{abstract}
		
		
	\end{frontmatter}
	
	%\linenumbers
	\tableofcontents
	\section{Introduction}
Over the course of less than a century, polymers have become pervasive in everyday life and high-technology \cite{peacock2012polymer,hiemenz2007polymer,polymer-electronic,HUAN2016236,qitan,mayer2007polymer,haque2020synthesis,leigh2020helical,polymer-gas-separation,sequeira2010polymer}. Mass production of niche polymers, such as polyethylene, polypropylene and polystyrene, has outstripped the production scale of iron and steel, which have been the staple materials for millennia \cite{geyer2017production}. Different parts of the practically infinite chemical space of polymers display a dizzying variety of distinctive properties, which can be tuned exquisitely through control of their chemical and morphological structure \cite{peacock2012polymer,hiemenz2007polymer}. Extensive efforts have been devoted to searching the chemical space and tinkering with their structure and chemistry to optimize their properties for specific applications. Traditional intuition-driven and/or trial-and-error approaches have already revealed the promise that the polymer class of materials holds. Nevertheless, given the vastness of the chemical and structural space, new methods are required to effectively and efficiently search this space to identify optimal, application-specific solutions.

The field of polymer informatics attempts to address this daunting search problem by the utilization of modern data- and information-centric approaches, inspired by emerging artificial intelligence (AI) and machine learning (ML) methods \cite{ml5,AI1,ml2,ml3,ml4,oweida_mahmood_manning_rigin_yingling_2020,Kononova2019,ramprasad2017machine}. Polymer informatics efforts are seeing heightened activity and successes in recent years \cite{ramprasad2017machine,Debra-Polymer,soft-informatics,pg,pg2,chandrasekaran2020polymer,MANNODIKANAKKITHODI2017}, but many of the ideas and concepts have gradually taken shape over a period of decades \cite{ramprasad2017machine,Debra-Polymer,soft-informatics,chandrasekaran2020polymer,adams2008engineering}.

\begin{figure}[h]
\begin{center}
\includegraphics[width=\textwidth]{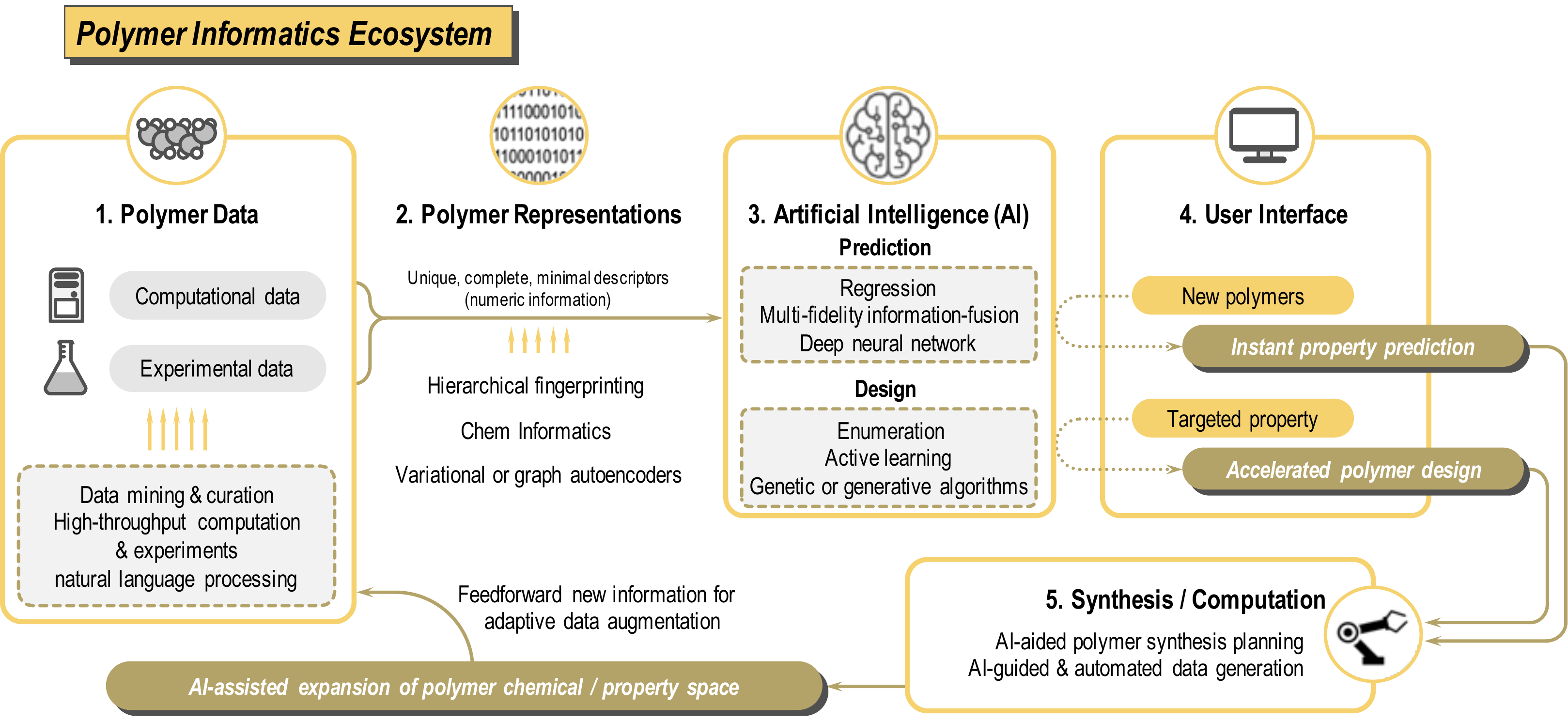}
\caption{Essential elements of Polymer Informatics Ecosystem: 1) polymer data, derived from (high-throughput) computations and/or experiments (through manual or natural language processing-aided excerption); 2) polymer representations,  transforming polymers into numerical fingerprints and making it amenable to ML/AI models; 3) developing surrogate models for polymer property prediction and design polymers with desired properties for specific applications; 4) Online user interfaces provide easy and quick public access to the developed surrogate models and/or the underlying polymer data; 5) AI-aided computational and synthesis validation, feeding new information to existing polymer repositories. }\label{polymerinformatics}
\end{center}
\end{figure}

  Figure  \ref{polymerinformatics} illustrates the essential elements of polymer informatics. The first vital ingredient is the polymer data, derived from experiments and (high-throughput) computations. Unlike hard materials, only limited well-organized/clean polymer data is available to be used for ML or AI-based techniques, e.g., in polymer handbooks \cite{polymer-data-book} and online repositories \cite{polyinfo}. Large volumes of experimental data remain trapped in the scientific literature, which is occasionally mined via laborious manual excerption.  An emerging alternative approach is natural language processing (NLP) to continuously and dynamically extract polymer data, but significant future efforts are needed to effectively and accurately extract polymer data from literature.  Another important resource of polymer data is high-throughput computations using density functional theory (DFT) \cite{Huan:Data,chen-dielectric,chen-esw} and classical molecular dynamics (MD) simulations \cite{MD_1,ionic-conductivity,MD_deformation,MD_transport,Seo-hall,SPE-ML,Sanket-1,Sanket-2}.  The recent development of autonomous computational agents, composed of machine learning modules and high-throughput computations, holds great promise for polymer data generation \cite{Huan-JPCL}.

The second important component of polymer informatics is a suitable framework to create machine-readable polymer representations. Linear notations are commonly adopted to describe the chemical information of polymers, for instance, using Simplified Molecular-Input Line-Entry System (SMILES) \cite{smiles}. With SMILES as input, polymers are either directly fingerprinted using hierarchical polymer fingerprints \cite{pg,pg2} or molecular fingerprints \cite{kumar,Wu-thermalconductivity} that are widely used in cheminformatics.  Alternatively, optimal fingerprint representation (or latent knowledge) of polymers can be obtained using variational \cite{batra2020polymer} or graph autoencoders \cite{jin2020hierarchical,jin2018learning}. Designing fingerprints that fully capture not just the chemical and morphological information of polymers, but also how they were synthesized and processed is one of the most challenging parts of polymer informatics.  

Using the numerical polymer fingerprints and target property data as input, we move to the third part of the polymer informatics: polymer property prediction and design. In the former, various machine learning algorithms, e.g., non-linear regression \cite{pg}, multi-fidelity information fusion \cite{MF,MF2} and deep neural networks \cite{anand-solvent,Wu-thermalconductivity}, can be applied to learn the relationship between polymer fingerprints and their target property, resulting in a surrogate property prediction model for polymers. The developed surrogate model can instantly predict various properties of new polymers defined by the user.  Another key benefit of the polymer informatics ecosystem is to accelerate the discovery of polymers with target properties for various applications. Several polymer design algorithms have been proposed, e.g., screening candidates based on the ML predicted properties from a huge list of enumerated polymers, iteratively selecting the next interesting polymer using active learning, and producing desired hypothetical polymers using genetic or generative deep learning algorithms. These design approaches have significantly accelerated the polymer design process for capacitors \cite{Arun:review,Arun:design,chen2020frequency}, membrane separation \cite{kumar}, organic solar cells \cite{wu2020machine}, among others.  

Once the desired polymer candidates are proposed, the next step is to validate the polymers via computational methods and physical synthesis.  The former is manageable using AI-automated data generation agents that control computational workflows  (but are applicable to only those properties that are accessible through computations). The latter is a challenge, as the synthesis of the selected polymer candidates is not straightforward. Chemical reactions, precursors, reagents and processing conditions (temperature, pressure and solvents) must be identified for each polymer to successfully synthesize them. Attempts are being made to expedite this process by using AI-assisted synthesis planning and robotic/autonomous (retro)synthesis. Although computer-aided synthesis design for molecules was recently accomplished \cite{coley2017prediction,autonomous2}, there remains lots of scope and challenges for polymer synthesis planning, which is expected to blossom rapidly in the next several years. Moreover, the data obtained from these synthesized polymers and/or from their computations can be added into existing polymer repositories to re-optimize ML models and re-design or re-imagine the next experiments.

In this paper, we review these emergent components of the polymer informatics ecosystem and discuss imminent challenges and opportunities. In Section \ref{data}, we discuss protocols available for polymer data generation, acquisition and management. It is followed by a survey of various schemas for polymer representations in Section \ref{feature}. Next, we move on to review machine learning algorithms utilized and adapted for polymer property prediction (Section \ref{predictionmodel}) and design for various applications (Section \ref{polymerdesign}). We then list several representative application examples that have benefited and may benefit from the polymer informatics philosophy in Section \ref{applications} and identify critical next steps that the community will need to address and surmount in the near future in Section \ref{nextstep}.

\section{Data generation, acquisition and management}\label{data}

The central tenet of polymer informatics is that if a sufficient volume of polymer data can be appropriately generated or curated, it can facilitate discovery/design of functional polymers with targeted performance. Below we discuss how polymer data can be accumulated from the literature or generated using high-throughput and autonomous computations.

\subsection{Scientific literature}

A reliable and enormous data resource for polymer data is the scientific literature, including printed handbooks \cite{brandrup1989polymer,wypych2016handbook,van2009properties,Mark2009PolymerDH}, online repositories \cite{polyinfo} and journal articles. As listed in Table \ref{database}, polymer handbooks, such as the Polymer Handbook \cite{brandrup1989polymer} and Properties of Polymers \cite{van2009properties}, are  introductory materials containing chemical, property and synthesis information on polymers. More recently, several polymer databases have been digitalized, allowing for easy access to polymer data. A few representative databases include PoLyInfo supported by the National Institute for Materials Science of Japan (NIMS) \cite{polyinfo}, CROW Polymer Property Database \cite{crow}, Polymers: A property database 
\cite{ellis2008polymers}, CAMPUS \cite{campus}, LANDOLT-BORNSTEIN \cite{LANDOLT-BORNSTEIN} and Polymer Property Predictor and Database (NIST) \cite{pppd}. In contrast to the field of inorganic materials, only a few computation-based property databases for polymers are available. This can be attributed to the high computational complexity of polymers due to their complicated physical and chemical structures. A good example of a database of computational data  polymers is Khazana \cite{khazana}, which includes DFT computed band gap, dielectric constant, refractive index and charge injection barriers. The third important resource of polymer data is the ever-increasing corpus of published journal articles.

\begin{table}[htp]
\caption{Available polymer data resources}
\small
\begin{tabular}{llll}
\hline
\hline
Source   & Name         & Data type  \\
\hline
Handbook&Polymer Handbook \cite{brandrup1989polymer}, Handbook of Polymers \cite{wypych2016handbook},  & Empirical\\
&Properties of Polymers \cite{van2009properties}, Polymer Data Handbook \cite{Mark2009PolymerDH}  & Empirical\\
&Polymer synthesis: theory and practice\cite{braun2012polymer}&Empirical\\

\hline
Online Repositories&PoLyInfo \cite{polyinfo}                        & Empirical    \\
&CROW Polymer Property Database  \cite{crow}                  & Empirical   &    \\
&Polymers: A property database \cite{ellis2008polymers}             & Empirical   &    \\
&CAMPUS \cite{campus}                                         & Empirical   &  \\
&LANDOLT-BORNSTEIN \cite{LANDOLT-BORNSTEIN}&Empirical   &  \\
&Polymer Property Predictor and Database (NIST) \cite{pppd} & Empirical   &  \\
&Khazana    \cite{khazana}                                             & Computational  &  \\
\hline
Published journal articles&Various&Empirical/Computational\\
\hline
\hline
\end{tabular}
\label{database}
\end{table}

Timely dynamical extraction of polymer data from the literature in a machine-readable format can be challenging and is achieved using either the laborious manual excerption and/or machine-learning methods usually classified as NLP. Manual text excerption refers to the old-fashioned procedure of collecting data from the literature and entails laborious efforts for data extraction, validation, and normalization owing to the absence of standard journal policies for publishing polymer data. Nonetheless, researchers have painstakingly collected important information on polymer types, their chemical structures (repeat units), names and class, their properties (e.g., physical, thermal, mechanical, dielectric, physicochemical and solution properties), and their synthesis recipes (e.g., polymerization paths, reactants and reagents) \cite{brandrup1989polymer,wypych2016handbook,van2009properties,Mark2009PolymerDH,polyinfo}. Crucially, easy access to the resulting databases has been provided through online repositories, as discussed earlier.

Machine learning-based NLP has emerged as an alternative approach for information excerption from the literature in the last few years. NLP  can be used to automatically scan the literature corpus and extract relevant polymer properties, which can be organized in a tabular fashion based on the NLP model predicted text relations. The use of NLP in materials science is still in its infancy, due to the difficulty in interpreting technical languages and incorporating domain knowledge. It is further complicated by the absence of standard journal policies for publishing scientific data.  Several initial attempts have tried to use NLP to collect materials synthesis recipes, capture latent knowledge and to even predict potentially superior thermoelectrics \cite{tshitoyan2019unsupervised,Kim2017,Kononova2019}. These successes motivate the further application of NLP in polymer informatics, such as the extraction of  properties, synthesis recipes or processing conditions from past literature. Despite initial success,  many unique challenges are posed in the case of polymers, for example, non-uniform polymer names. More details are described in Section \ref{nextstep}.

\subsection{High-throughput and autonomous computational agents}

High-throughput computations using first-principles theory and classical MD are important approaches to amass polymer data. However,  this task is non-trivial because of the enormously complicated chemical and physical structures of polymers at the atomic scale; polymers usually display either amorphous or semi-crystalline phases. Given the expensive computational cost of first-principles computations, small length-scale models ($<$ 100 atoms) have been developed to represent polymers and approximate their physical, electronic and dielectric properties \cite{Huan:Data,Arun:design,Huan-JPCL,chen-esw,chen-dielectric}. The computed results, however, generally suffer from certain accuracy issues depending on the methodology. To model polymers in the large-scale,  classical MD with empirical force fields have been applied to study the structural dynamics \cite{yi2013molecular,MD_crystallization,MD_Yinling_2,MD_morphology,Lihua:SR,ReaxFF_1,ReaxFF_2,ReaxFF_3} and diverse properties (e.g., dielectric, thermal, mechanical and ion transport properties)\cite{fukushima2019effects, thermal-conductivity-1,thermal-conductivity-2,mechanical-property-1,MD_1,MD_deformation,MD_tensile,MD_mechanical,MD_Yinling_2,SPE-ML,MD_transport,Sanket-1} of polymers.  However, such parameterization schemes are also restricted by the availability of force fields and the high computational cost to simulate very large systems ($>$ thousands of atoms). 

\begin{figure}[hbt]
\begin{center}
\includegraphics[width=1.0\textwidth]{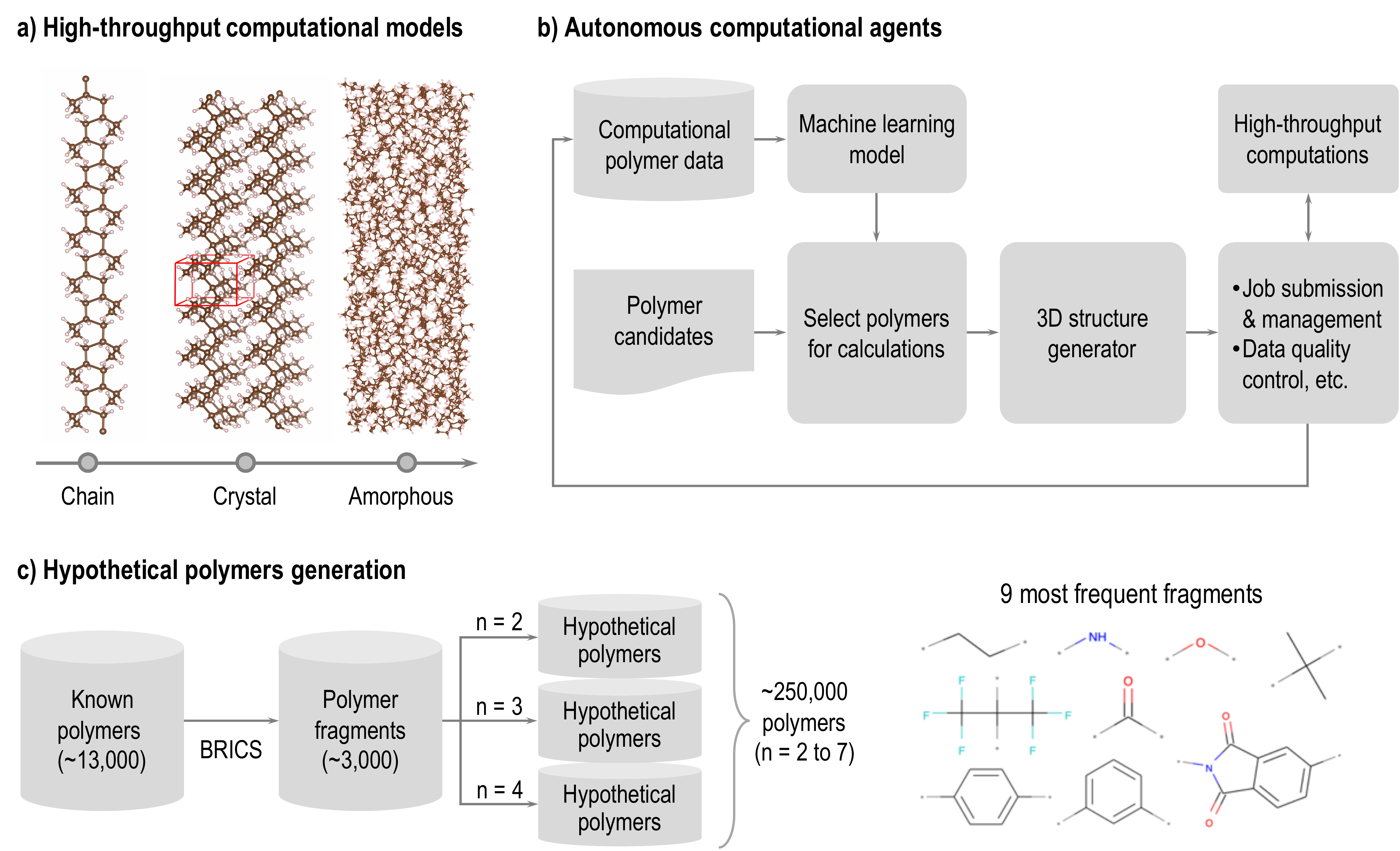}
\caption{a) Hierarchy of models to represent polymers, i.e., single-chain, pure crystal and amorphous phases. This figure is taken from Ref. \cite{Huan-JPCL} with permission from ACS Publications. b) Autonomous computational agents to generate polymer data. c) Design of hypothetical polymers using the BRICS scheme, along with some common polymer building blocks. `*' represents the possible linking position for each building block. }\label{fig:data-generation}
\end{center}
\end{figure}

Balancing the trade-off between cost and accuracy, past efforts have led to the computation of the electronic, dielectric and optical properties (such as band gap, charge injection barriers and dielectric constant) of thousands of polymers using DFT \cite{Huan-JPCL,Huan:Data, Kamal_2020,chen-dielectric,chen-esw}. A hierarchy of models, i.e., single-chain, pure crystal and amorphous, has been developed to represent realistic polymers, as shown in Figure  \ref{fig:data-generation}a). The simplest single-chain model is composed of only a chain of monomers in vacuum, while crystal and amorphous models represent the crystalline and  amorphous regions of polymers, respectively. In spite of this simplification, the creation of correct low-energy crystalline structure of a polymer, especially for novel polymers, remains a major challenge \cite{Huan-JPCL}. To address this issue, Huan \textit{et al.} developed a general computational workflow, referred to as polymer structure predictor (PSP), to predict crystal structures of linear polymers. In this workflow, a polymer is defined in terms of its chemical composition and atomic connectivity, using the SMILES notation (more details in Section \ref{feature}). Reasonable single-chain and crystal models of the polymer can be predicted/created \cite{Huan-JPCL} using this scheme. Such efforts have led to formation of the largest dataset for polymers using computations, which can be accessed from https://khazana.gatech.edu. Some of the important computed properties include the crystal band gap, single-chain band gap, charge injection barriers, atomization energy, ionization energy, electron affinity, dielectric constant and refractive index \cite{Huan:Data,Huan-JPCL}. Other researchers have spent extensive efforts on estimating  thermal conductivity \cite{thermal-conductivity-1,thermal-conductivity-2,thermal-conductivity-3,thermal-conductivity-4}, Young’s modulus \cite{MD_mechanical}, tensile strength \cite{mechanical-property-1,MD_deformation,MD_tensile}, and the lithium conductivity \cite{SPE-ML,ionic-conductivity,Seo-hall} of representative polymers using classical MD simulations. However, it is still challenging to compute these properties for a diverse range of polymers, especially those that have not been studied well.

Given the vast chemical space of polymers,  a new strategy aided by an autonomous computational agent has been developed to dynamically select the next-candidate polymer with target properties \cite{Huan-JPCL}. As visualized in Figure  \ref{fig:data-generation}b), by utilizing the available (seed) computational dataset, machine-learning models are developed with the capability to instantly predict target properties for a large number of new polymers. Candidate polymers that meet the desired properties are selected, followed by a ``3D structure" conversion step (involving generation of hierarchical models as shown in Figure \ref{fig:data-generation}a)). The newly created structures are then modeled using high-throughput calculations, and the obtained results are added to the seed dataset iteratively.  This autonomous platform is applicable for single or multiple polymer property predictions, and offers an efficient way to discover/design polymers with desired performances.

\subsection{Hypothetical polymers}

Data derived from experiments or the computations mentioned involve only known polymers.  But how can we expand and explore beyond the \emph{known} polymer chemical space? Variations of this question have already been tackled by different communities within chemical and materials sciences, such as drug discovery, inorganic solid state, metal-organic frameworks, 2D materials, etc., by exploiting various theoretical tools to construct databases of hypothetical molecules (e.g. ZINC) or materials (e.g. Materials Project). Further, computational tools, such as first-principles or classical methods, have been employed to estimate properties of these hypothetical cases, and screen promising candidates for future synthesis efforts. Some databases even estimate the synthesizability of a candidate using multiple models (e.g., based on free energy, synthetic accessibility scores, etc.) to ensure only realistic and plausible candidates are included. The successes of such molecule/materials databases are inspiring the creation of similar libraries for polymers.

In this regard, Batra \textit{et al.} \cite{batra2020polymer} devised an approach to explore the vast \emph{unknown}  chemical polymer space by generating new, but realistic, hypothetical polymers. As illustrated in Figure  \ref{fig:data-generation}c), they first obtained SMILES representation of $\sim$ 12,000 polymers successfully synthesized in the past. Next, using the concepts of breaking of retrosynthetically interesting chemical substructures (BRICS) \cite{BRICS, RDKit}, polymer ``building blocks'' --- each with two or more chain ends denoted by symbol \texttt{[*]}, e.g., \texttt{[*]c1ccc([*])cc1}, \texttt{[*]C(=O)[*]}, \texttt{[*]CC[*]} --- were obtained along with their frequency of occurrence. Following this, hypothetical polymer SMILES strings were created by combining (at the [*] location) various numbers of building blocks, ranging from 2 -- 7, resulting in a total of $\sim$ 250,000 hypothetical polymers.  This approach can be extended to create nearly infinite polymer candidates that may later be used to screen target properties. Care is taken to preserve the frequency of occurrence of different building blocks, making the constructed hypothetical SMILES dataset realistic and representative of the initial collected empirically known polymers. Moving forward, different chemical constraints or block neighbor restrictions can be introduced to limit the possible combination of building blocks, and generate more realistic/synthesizable polymers.

\section{Polymer representation}\label{feature}

Once the polymer structural, chemical, property and synthesis data are collected from the aforementioned resources, it should be processed/transformed to make it amenable to AI/ML based methods. Depending on the target polymer property or the input data type, different polymer representations may be chosen. Below, following a short discussion on the group contribution method, we discuss some of the more recent and successful polymer representation methods.

The group contribution technique developed by Van Krevelan and coworkers assumes that a polymer property can be estimated as a weighted sum of contributions arising out of its constituting fragments (referred to as quantitative structure-property relationships (QSPR) fingerprints) \cite{property-prediction}. Using this and subsequent variations of this method, models describing a range of polymer properties have been developed with/without machine learning, including glass transition temperature, dielectric constant, refractive index, electrical conductivity, thermal conductivity, gas and aqueous diffusion, and intrinsic viscosity \cite{property-prediction, QSPR-Tu,yu2006prediction,permea-NN}. However, the developed models rely on the available fragment library and have little predictive capabilities for new polymers containing chemical fragments outside this pre-defined library. 

\begin{figure*}[h]
	\centering
	\includegraphics[width=1.0\textwidth]{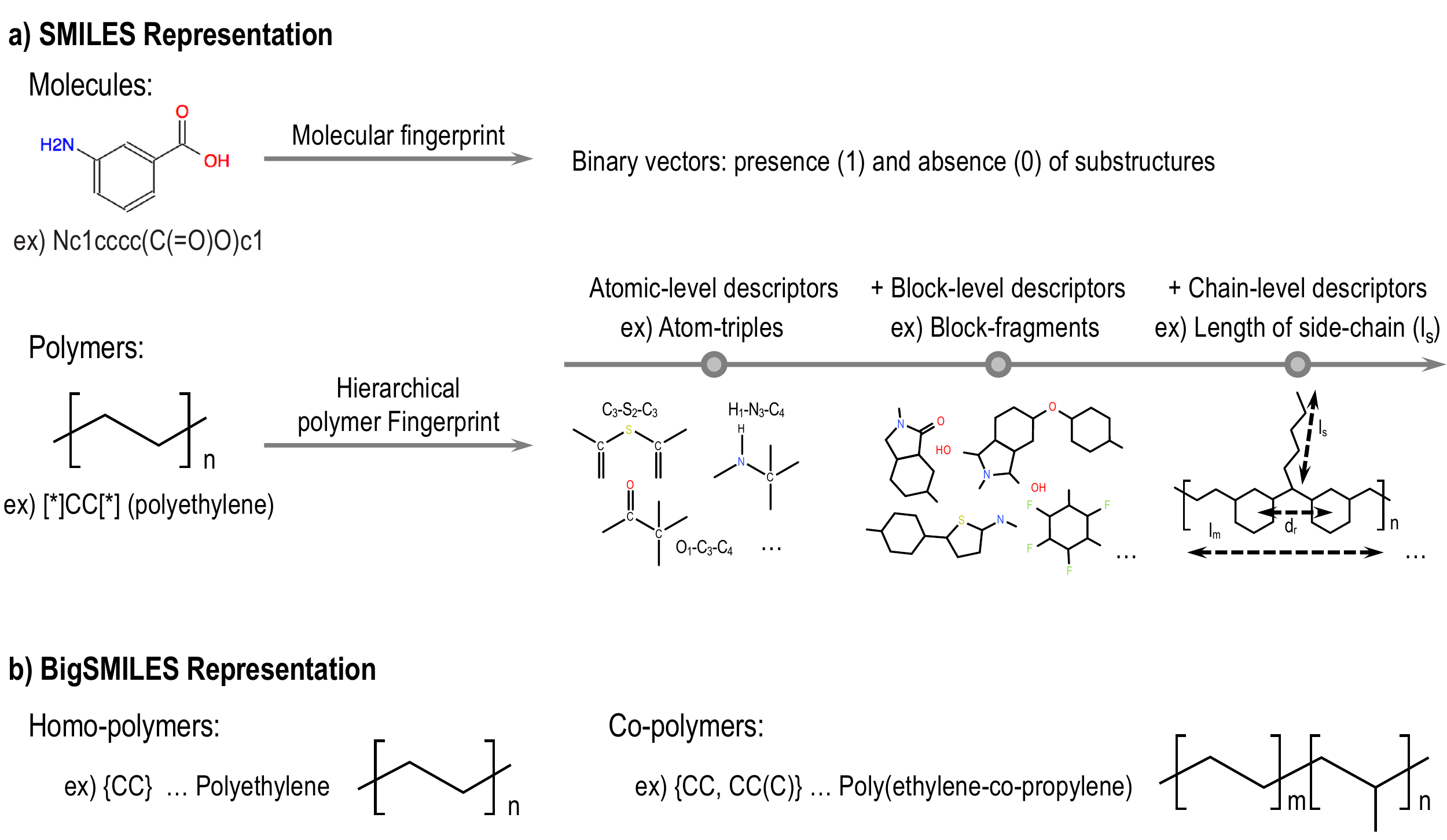}
	\caption{Polymer representations: SMILES and BigSMILES \cite{bigsmile}. Using the input of SMILES, molecular \cite{kumar} and hierarchical polymers fingerprints \cite{pg,pg2} were developed to numerically represent polymers.}
	\label{reprentations}
\end{figure*}

To efficiently encode chemical information of molecules into machine-readable format, line notations have been designed to describe molecules using a line of text strings. Examples of such approaches include SMILES, the Wiswesser line notation (WLN), SYBYL Line Notation (SLN) and IUPAC International Chemical Identifier (InChi). SMILES is one of the most popular methods to represent molecules, because it is both human-readable and machine-friendly \cite{smiles}.  Further, various molecular fingerprinting algorithms have been developed to transform SMILES of small molecules into numerical vectors. Avalon, Daylight and Extended-Connectivity \cite{ECF} fingerprints are examples of common fingerprinting algorithms that can be accessed through the open-source RDkit library \cite{RDKit}. Within these fingerprints, the presence or absence of substructures within a molecule is encoded into binary vectors, which can be used as inputs to data-driven models.  SMILES representations of molecules can also be utilized (or transformed as graphs) in generative neural networks for fingerprint learning and molecular generations \cite{jin2018learning,jin2018learning}, but can also be directly used as input language in text-based machine learning algorithms \cite{segler2018generating,goh2017smiles2vec}. The use of SMILES and similar line notations for molecules has transformed data-driven research in chem- and bio-informatics.

In contrast to small molecules, polymers are macromolecules composed of many repeat units, and require unique ways to capture their structural information.  As illustrated in Figure \ref{reprentations}a), in modern data-driven models, SMILES of oligomers with several repeat units ($<5$) have been applied to represent polymers, which can be fingerprinted using regular molecular fingerprinting algorithms \cite{Wu-thermalconductivity,kumar,thermal-conductivity-4}. The ML models developed using such oligomer fingerprints can predict various properties of polymers fairly well, although the effect of polymer morphology on the target property is excluded. In a similar development, modified SMILES representations for polymers have been developed which represent endpoints or connection points of repeat units using special symbols. For example, polyethylene is represented as [*]CC[*], where CC is the repeat unit of polyethylene and * represents the connecting points between repeat units \cite{pg,pg2}. Furthermore, a hierarchy of hand-crafted fingerprints for polymers have been developed to capture the connectivity and morphology information of polymers in order to  improve the property prediction accuracy \cite{Wang2014979,wu2016prediction,mannodi2016machine,mannodi2016rational,pg}. Figure \ref{reprentations}a) shows details of the hierarchical fingerprint, including the (1) atomic-level, (2) block-level, and (3) chain-level components. At the atomic-level, the number fraction of atomic-level fragments within the polymers, defined by the generic label ``A$_i$B$_j$C$_k$'', are considered.  The block-level fingerprint components are the number fraction of pre-defined building blocks that constitute the polymers, such as C$_{6}$H$_{4}$, C=O, CH$_{2}$ and CO. Chain-level features capture information at the morphological scale, including the length of the longest or shortest side chains with or without rings. Further, QSPR fingerprints, such as the volume to surfaces ratio and van der Waals surface area, are also considered. Using this approach, models to predict many properties of polymers have been developed, including band gap, glass transition temperature and dielectric constant \cite{pg,pg2}. Detailed examples are provided in Section \ref{predictionmodel}. 

Additionally, BigSMILES has been recently developed for describing macromolecules, e.g., homo-and co-polymers \cite{bigsmile}. It is an extension to SMILES, expressed as $\{$RepUnit1, RepUnit2, RepUnit3, …$\}$. Here, RepUnit1, RepUnit2, RepUnit3 are a list of (same or different) repeat units within polymers, with random positions. For example, Poly(ethylene-co-propylene) is denoted by $\{$CC, CC(C)$\}$, as shown in Figure \ref{reprentations}b). In this representation, branched, network and terminal group information of polymers may be  also incorporated. However, there are no available fingerprinting algorithms to transform BigSMILES into numerical vectors yet.

Molecular structures can also be represented as a graph via an input of SMILES,  where atoms and bonds are represented by nodes and edges, respectively. Such a method has been widely utilized for molecular structure generation and property prediction in cheminformatics, bioinformatics and materials science with great success \cite{GNN1,GNN2,GNN3,GNN4,zeng2018graph}. Since it is challenging to use graphs to represent polymers, due to its large-scale morphology, researchers have attempted to use oligomers (including less than 5 repeat units) to label polymer graphs with atom-based or substructure/motif based-methods \cite{jin2017predicting,jin2018learning,jin2020hierarchical}. Motifs refer to larger size substructures. However, because monomers of many polymers are large and complicated, this leads to monomer generation failures using small substructures.  Further, polymers are composed of large numbers of monomers, and it is not clear how one can incorporate large-scale morphological information as graphs. These critical issues are discussed in Section \ref{nextstep}.

\begin{figure*}[t]
	\centering
	\includegraphics[width=1\textwidth]{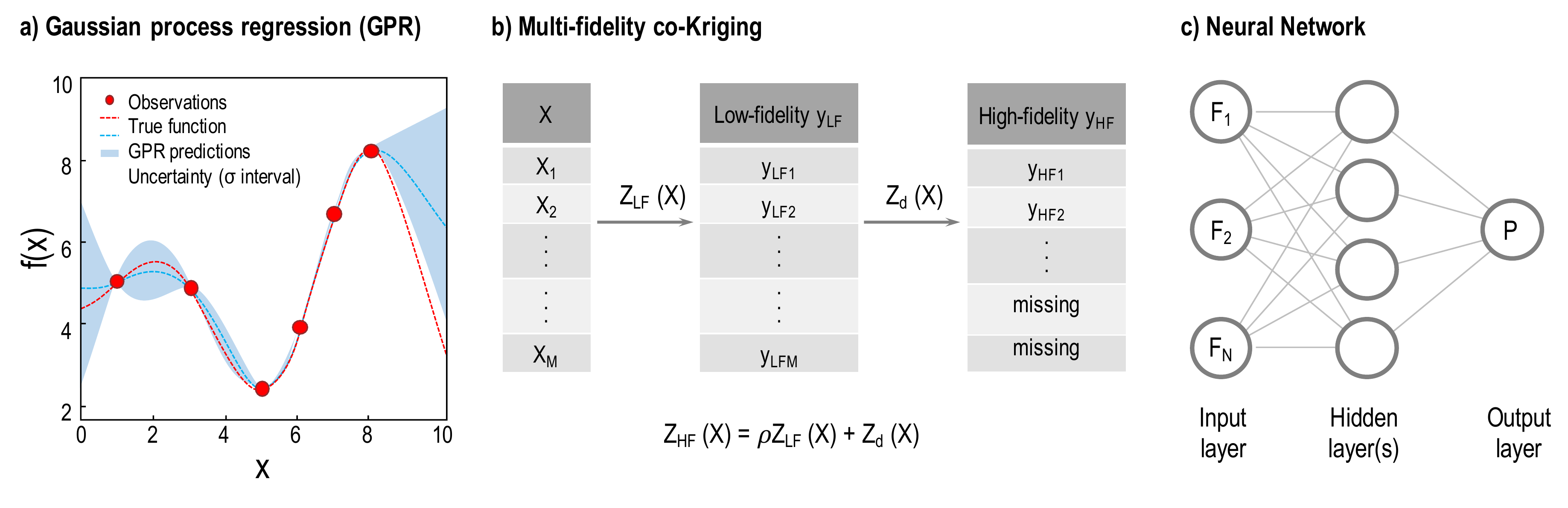}
	\caption{a) Gaussian process regression (GPR) model to learn the correlation between fingerprints and target property, providing predicted values and uncertainty (shaded regions). b) Multi-fidelity (MF) co-kriging approach depends on two models: the Gaussian processes $Z_{\rm LF}(\boldsymbol{x})$ of the low-fidelity (LF) function mapping the fingerprint space and low-fidelity property (y$_{\rm LF}$) and the Gaussian processes $Z_{\rm d}(\boldsymbol{x})$ to map the fingerprint space and difference between the low-fidelity and the high-fidelity (HF) functions. The property prediction at the high-fidelity level ($Z_{\rm HF}(\boldsymbol{x})$) is $Z_{\rm HF}(\boldsymbol{x}) = \rho Z_{\rm LF}(\boldsymbol{x}) + Z_{\rm d}(\boldsymbol{x})$, where $\rho$ is a scaling factor that quantifies the correlation between the two fidelities of data.
c) General Neural network workflow, including input, hidden and output layers.}
	\label{ML}
\end{figure*}

\section{Property prediction schemes}\label{predictionmodel}

The selection of suitable learning algorithms to map polymer fingerprints and properties is a critical step. Depending on the complexity of the target property, the volume and the nature of the available datasets, various learning algorithms have been applied, such as linear or non-linear regression algorithms,  multi-fidelity information fusion and deep neural networks.

\subsection{Linear/Non-linear regression}

	\begin{figure*}
		\centering
		\includegraphics[width=1.0\textwidth]{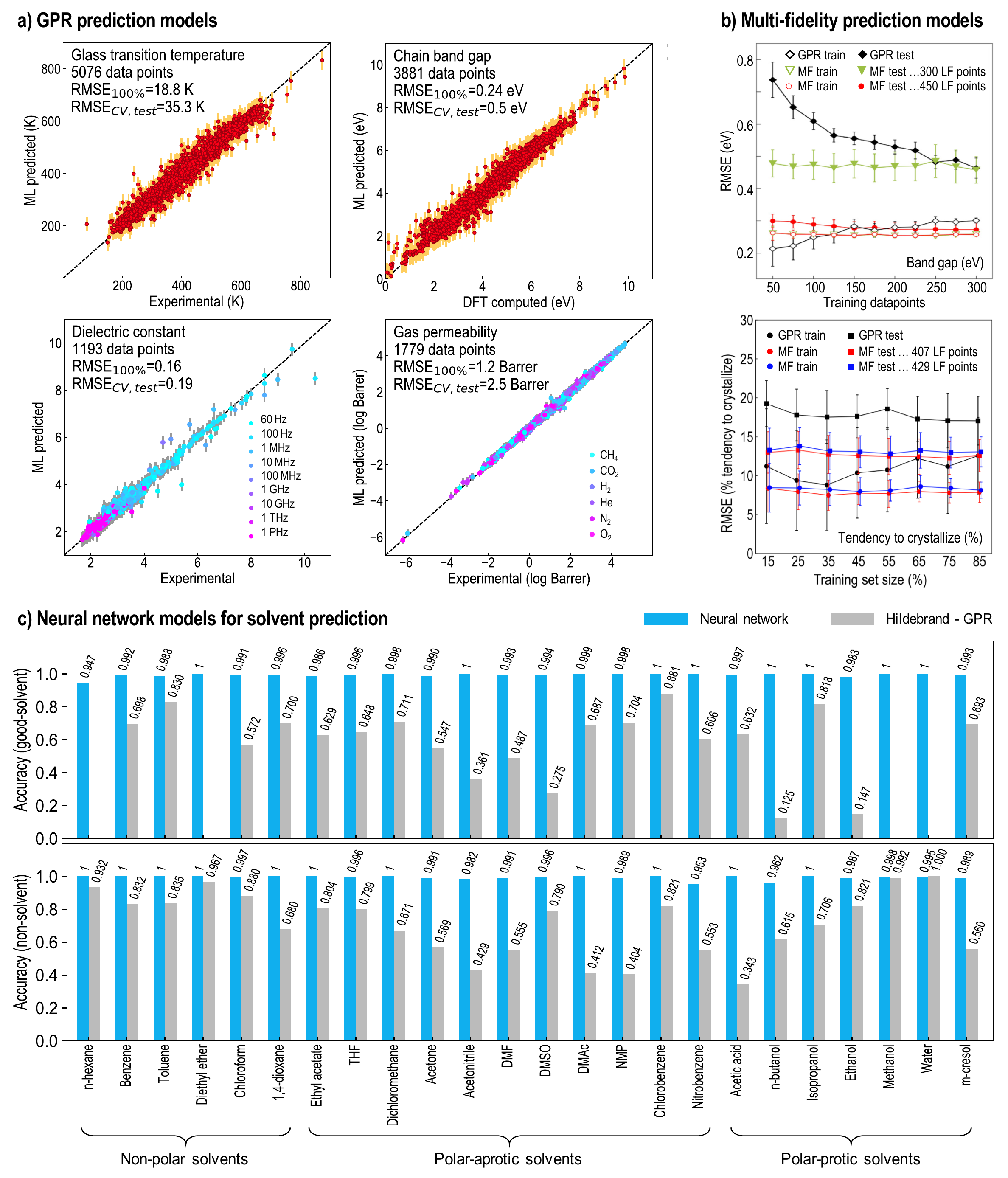}
		\caption{a) Parity plots of GPR predicted and true values of the glass transition temperature, band gap of sing-chain polymers, dielectric constant and gas permeability. In the case of the gas permeability model, 6 gases, i.e., O$_2$, N$_2$, CH$_4$, He, CO$_2$ and H$_2$, were considered and dielectric constant at 9 different frequencies was used in the dielectric constant model. CV-test RMSE is the average RMSE of the test subsets in 5 fold-CV \cite{pg2}. Error bars represent GPR uncertainty. b) A comparison of learning-curves for the GPR and multi-fidelity (MF) predicted band gap \cite{MF2} and tendency to crystallize \cite{Shruti-crystalinity}. 
		c) Neural network-based solvent prediction accuracy of soluble (top) and insoluble (bottom) polymers for 24 solvents, including non-polar, polar-aprotic and polar-protic solvents, along with results from GPR models trained by Hildebrand parameters \cite{anand-solvent}. Figure a), b) and c) are taken from Ref. \cite{pg2}, Ref. \cite{MF2,Shruti-crystalinity} and Ref. \cite{anand-solvent}, respectively,  with permission from ELSEVIER and ACS Publications.}
		\label{model}
	\end{figure*}

The linear regression algorithm assumes that the property being modeled is a linear function of the fingerprints, which is the simplest method to build machine learning models. For polymers, various property prediction models have been developed using group contribution methods \cite{property-prediction}, multiple linear regression \cite{QSPR-Tu,yu2006prediction}, and least-squares regression \cite{soft-informatics,jabeen2017refractive,venkatraman2018designing}, with QSPR or quantum-chemical fingerprints. These models are limited by the neglect of the non-linear relationships between polymer fingerprints and their properties. To overcome this issue,  non-linear regression algorithms have been employed,  such as support vector machine (SVM), kernel ridge regression (KRR) and Gaussian process regression (GPR). For instance, Yu \textit{et al.} used SVM to train glass transition prediction models using QSPR fingerprints and experimental property values \cite{QSPR-Tu,yu2010support}, while KRR has been applied to develop models for a series of high-throughput computed polymer properties (e.g., band gap and dielectric constant)\cite{pilania2013accelerating}.

In recent years, the GPR algorithm has been broadly utilized to build machine learning models for polymer property prediction \cite{pg,pg2,zhu2020polymer,kumar,chen2020frequency}. As illustrated in Figure  \ref{ML}a), the key advantage of GPR is that predicted uncertainties are provided by learning a generative and probabilistic distribution with the mean representing the prediction and the confidence interval estimating the uncertainty.  Figure  \ref{model}a) illustrates four representative GPR models, including chain band gap, glass transition temperature,  frequency-dependent dielectric constant and gas permeability. These models were trained using 3881 DFT computed, and 5076, 1193 and 1779 experimental values, respectively. 5-fold cross-validation (CV) was employed to avoid model overfitting. R$^2$ and RMSE denote the coefficient of determination and the root-mean-square error, respectively. RMSE$_{100\%}$ and RMSE$_{\rm CV,test}$ respectively denote the RMSE errors on the whole dataset used for model training or  on the test subset during 5 fold-CV. In the case of the gas permeability model, 6 gases, i.e., O$_2$, N$_2$, CH$_4$, He, CO$_2$ and H$_2$, were considered and numerically represented using one-hot encoding. Likewise, in the dielectric constant model the frequency value (at 9 different frequencies) was used as a feature to obtain a frequency-dependent dielectric constant model. We note that the developed GPR models exhibit very high R$^2$ and acceptable RMSE$_{\rm CV,test}$ with respect to the wide property range of the training datasets. Additionally, the performance of these models has been further validated by using systematic analysis, involving the effect of feature reduction, various levels of train-test splits (i.e., learning curves) and validation on completely unseen datasets.

The GPR algorithm can be used to build accurate and reliable ML models for a single property while also providing prediction uncertainties.   However, the GPR method has two issues: 1) it requires a manageable dataset size. Large datasets ($>$ 5000) become prohibitively expensive to train. 2) It does not have the capability to train multiple properties in one single model. Therefore, more advanced algorithms have been utilized to improve these issues, such as multi-fidelity information fusion and deep learning methods as discussed below.

%A popular kernel choice for GPR is the radial basis function (RBF) kernel, expressed as 
%\begin{equation}\label{eq:kernel}
%k(\boldsymbol{x},\boldsymbol{x'}) =  \sigma_{f}\exp \left( -\frac{1}{2\sigma^2_{l}}||\boldsymbol{x} -\boldsymbol{x'}  ||^2 \right) + \sigma^{2}_{n}.
%\end{equation}
%Here, three hyperparameters $\sigma_{f}$, $\sigma_{l}$ and $\sigma_{n}$ represent the variance, the length-scale parameter and the expected noise in the data, respectively. These are determined during the model training by maximizing the log-likelihood estimate which mathematically turns out to be the same as minimizing model error. Here, $\boldsymbol{x}$ and $\boldsymbol{x'} $ are the fingerprint vectors of the two polymers. 

\subsection{Multi-fidelity information fusion  approaches} 

It is quite common to encounter problems where several datasets have varying levels of accuracy, data generation cost and noise levels are present. Typically, the most precise experimental measurements (or computations) also tend to be the most time and resource intensive, the so-called high-fidelity (HF) data. However, polymer properties of interest can also be estimated via cheaper methods at lower accuracy. For instance, empirical trends, simple group-contribution methods and computationally demanding quantum mechanical simulations can generate this low-fidelity (LF) data. Given such a situation, a multi-fidelity (MF) information fusion model aims to consolidate all the available information from the varying fidelity sources to make the most accurate and confident property predictions at the highest level of fidelity \cite{pilania_data-based_2020, MF2, MF, zaspel_boosting_2018, pilania_multi-fidelity_2017, lee_prediction_2016, ramakrishnan_big_2015}. Comparative studies have shown that the multi-fidelity approach performs better than any single-fidelity based method in terms of prediction accuracy, especially for small (high-fidelity) data sets. Typical strategies for MF learning are discussed in Ref. \cite{MF}. Among them, the Gaussian processes-based co-kriging regression method \cite{kennedy_predicting_2000} is viewed as a powerful method and has been utilized to predict polymer properties, such as band gap and degree of crystallinity \cite{MF2,Venkatram2019}. As shown in Figure  \ref{ML}b), this MF approach is composed of two models: the Gaussian processes $Z_{\rm LF}(\boldsymbol{x})$ of the low-fidelity function and the Gaussian processes $Z_{\rm d}(\boldsymbol{x})$ related to the difference between the low-fidelity and the high-fidelity functions. The property prediction at the high-fidelity level ($Z_{\rm HF}(\boldsymbol{x})$) is $Z_{\rm HF}(\boldsymbol{x}) = \rho Z_{\rm LF}(\boldsymbol{x}) + Z_{\rm d}(\boldsymbol{x})$. Here, $\rho$ is a scaling factor that quantifies the correlation between the two fidelities of data.

Figure  \ref{model}b) shows two successful examples of MF approaches being applied to polymer property predictions \cite{MF2,Shruti-crystalinity}, i.e., the tendency to crystallize and the band gap. For the former, a MF model was trained using 107 high-fidelity data directly measured by experiments and 429 low-fidelity data estimated using a combination of experimental and group-contribution methods. In the latter, 382 hybrid and PBE computed band gap values were used as high- and low-fidelity data in the MF model. Figure  \ref{model}b) compares the learning performance of the MF models against single-fidelity GPR models trained on the respective high-fidelity polymer property data, i.e., the RMSE on the training and the test set as a function of the training size of the high-fidelity data. We note that MF models surpass the GPR model accuracy (trained on the high-fidelity data alone) at a much smaller fraction of the high-fidelity training data. This is mainly because the relatively large volume of the low-fidelity data, although somewhat inaccurate, allows the MF model to learn polymer property trends. These findings indicate that there are benefits to employing the MF approach, especially in situations wherein resource demanding high-fidelity experimental data can be combined with a large number of low-fidelity and inexpensive computational data.

	While the first proof-of-principle examples are just beginning to appear, MF models could have a considerable impact in the field of polymer informatics. It is worth pointing out that the accuracy of MF models depends on the ability of the shared subset of high- and low-fidelity data to learn the latent space of the two fidelities. Further, there is a necessity to improve upon the MF scheme. For instance, several levels of fidelity hierarchies can be present simultaneously in the polymer property datasets. The number of the co-kriging model parameters can significantly increase in such scenarios, leading to expensive computational cost. Consequently, advanced MF learning algorithms should be developed to speed up the learning process, particularly when several levels of fidelities are present in the polymer data sets.

\subsection{Deep neural networks}
Conventional machine learning techniques described above provide good property prediction accuracy. However, these methods are computationally efficient for systems with small dataset size only. Given the surge in the available computational/experimental data in materials science, deep neural networks (NN) are being increasingly utilized in polymer informatics. Figure  \ref{ML}c) presents the general architecture of NNs, in which molecular or polymer fingerprints form the input layer. The following hidden layers are constructed with a specific activation function, e.g., the parametrized rectified linear unit (PReLU). The final output layer of the NN consists of neuron(s) for target properties, also with a specific activation function depending on the problem at hand (classification or regression). The details of various types of NNs and their uses in materials science are well-reviewed in Refs. \cite{NN1,NN2,NN3}. Below we discuss the initial attempts to apply NNs for polymer properties prediction \cite{liu2009artificial,chen2008neural,sumpter1994neural,anand-solvent,transfer,Wu-thermalconductivity}.

The selection of suitable polymer-solvent pairs is a critical step for polymer synthesis. Chandrasekaran \textit{et al.} have developed a deep neural network model for solvent prediction \cite{anand-solvent}.  In this work, 4,595 polymers and 24 solvents, forming a total of 11,958 polymer + good-solvent pairs and 8,469 polymer + non-solvent pairs, were used to train a binary classification NN model (i.e., given a polymer-solvent pair the model predicts if it a good-solvent or non-solvent for that polymer). A multilayer perceptron with special architecture was used:  the first part of the NN composed of two input branches, one for the polymer descriptors generated using hierarchical fragment-based fingerprint described in Section \ref{feature} and the other for the solvent descriptors represented by one-hot encoding. In the second part, polymer and solvent latent space were concatenated into a single merged latent vector. Figure  \ref{model} c) shows the neural network prediction accuracy of soluble (top) and insoluble (bottom) polymers for 24 solvents, including non-polar, polar-aprotic and polar-protic solvents. Performance results for the GPR models trained using Hildebrand parameters of about 100 polymers \cite{Venkatram2019} are also compared. In general, the performance of the NN model greatly outperforms that of the GPR model, mainly due to the higher level of diversity in the training data. Further, the Hildebrand parameter is only an approximate empirical approach to distinguish good-solvent against non-solvents, based on the notion of ``like dissolves like". This deep neural network-based framework provides a more general, accurate, and efficient way to predict good-solvents vs. non-solvents for new polymers.

Additionally, researchers have applied NNs to build prediction models for glass transition temperature \cite{liu2009artificial,chen2008neural,sumpter1994neural}, polymer permeability to gases \cite{permea-NN} and thermal conductivity ($\kappa$) of polymers \cite{Wu-thermalconductivity,transfer,thermal-conductivity-4}.  For the glass transition temperature and polymer permeability, small datasets ($<=$ 150 polymers) were applied to train NNs,  raising concerns on the generality of obtained models for new cases. In the case of $\kappa$, Zhu \textit{et al.} have used classical MD computed $\kappa$ values of single-chain polymers and molecular fingerprints to train the NNs models.  One potential concern is that the computational uncertainty, arising from the model difference between the adopted single chain and realistic polymers, may introduce additional noise in the ML models \cite{thermal-conductivity-4}.   An alternative approach is to directly use experimental $\kappa$ to train the model, although, only sparse data is available. To overcome this issue, Wu \textit{et al.} \cite{Wu-thermalconductivity} has utilized the transfer learning approach to learn the $\kappa$ of polymers (58 data points), via training other properties of polymers with large data size (e.g., melting temperature, glass transition temperature and heat capacities).  The performance of the obtained model is better than those trained only on the thermal conductivity data, because the shared features between $\kappa$ and other properties and large training dataset are considered in the transfer learning algorithm. However, there are still challenges, as discussed in Section \ref{nextstep}.

	\begin{figure*}[h]
		\centering
		\includegraphics[width=1.0\textwidth]{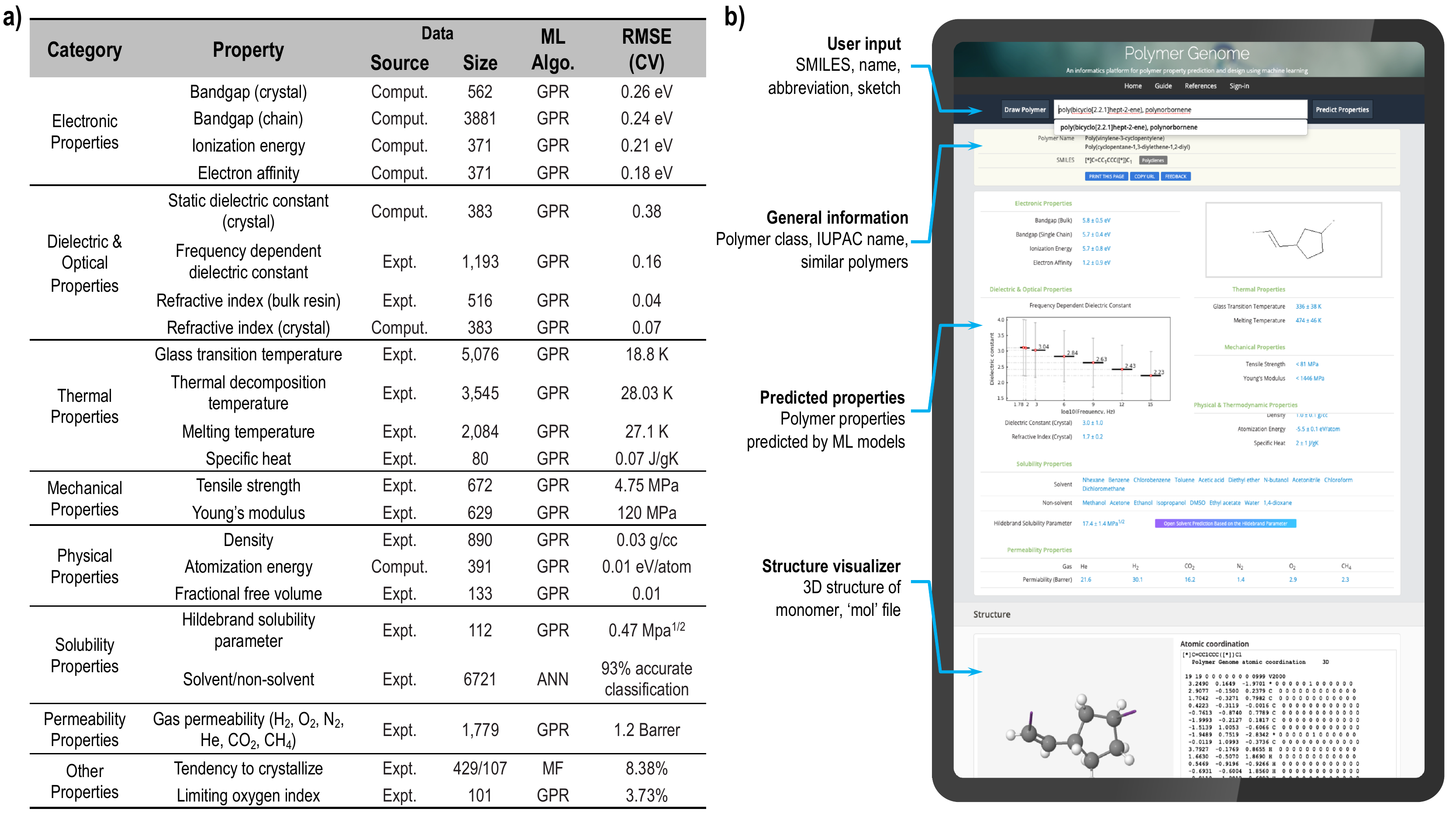}
		\caption{a) Summary of various property prediction models implemented in the Polymer Genome online platform (www.polymergenome.org). RMSE(CV) denotes the average RMSE errors on the test subset during the 5 fold-CV. b) Overview of the Polymer Genome platform. Polynorbornene is used as an example of user input to show the obtained ML predicted properties and 3D structure visualization. }
		%Figure a) is taken from Ref. \cite{pg2} with permission from AIP Publications.
		\label{pg}
	\end{figure*}

\subsection{Polymer Genome online platform}
Significant efforts are also being made to provide easy access to the aforementioned polymer prediction models. In this regard, the Polymer Genome online platform (www.polymergenome.org) has been developed to instantly provide property predictions for polymers. As summarized in Figure  \ref{pg}a), various polymer property prediction models have been implemented, including electronic, dielectric, thermal, mechanical and other important properties. The source and size of the training data, the applied algorithms and the expected errors (RMSE$_{\rm CV}$) for each of the property models are also provided.  Figure  \ref{pg}b) shows a typical output of Polymer Genome, taking the example of Polynorbornene. The SMILES ([*]C=CC1CCC([*]C1)), which forms the repeat unit of Polynorbornene, is provided as the input to Polymer Genome, where [*] denotes the connection points of the repeat units. ML predicted properties of this polymer are shown in a tabular format.  The 3D visualization of the structure with atomic coordinates is also provided at the bottom of the page. More detailed information is available in Ref. \cite{pg,pg2}.

\section{Polymer design algorithms}\label{polymerdesign}
Once the polymer surrogate models are trained, they can be utilized to accelerate the polymer discovery process. Below we outline two distinct strategies for this. While the first relies on screening candidates that meet target property requirements based on predictions for a pre-determined candidate pool, the other utilizes genetic and generative models to directly produce desirable candidates.

\subsection{Enumeration}

\begin{figure*}[hb]
		\centering
		\includegraphics[width=1.0\textwidth]{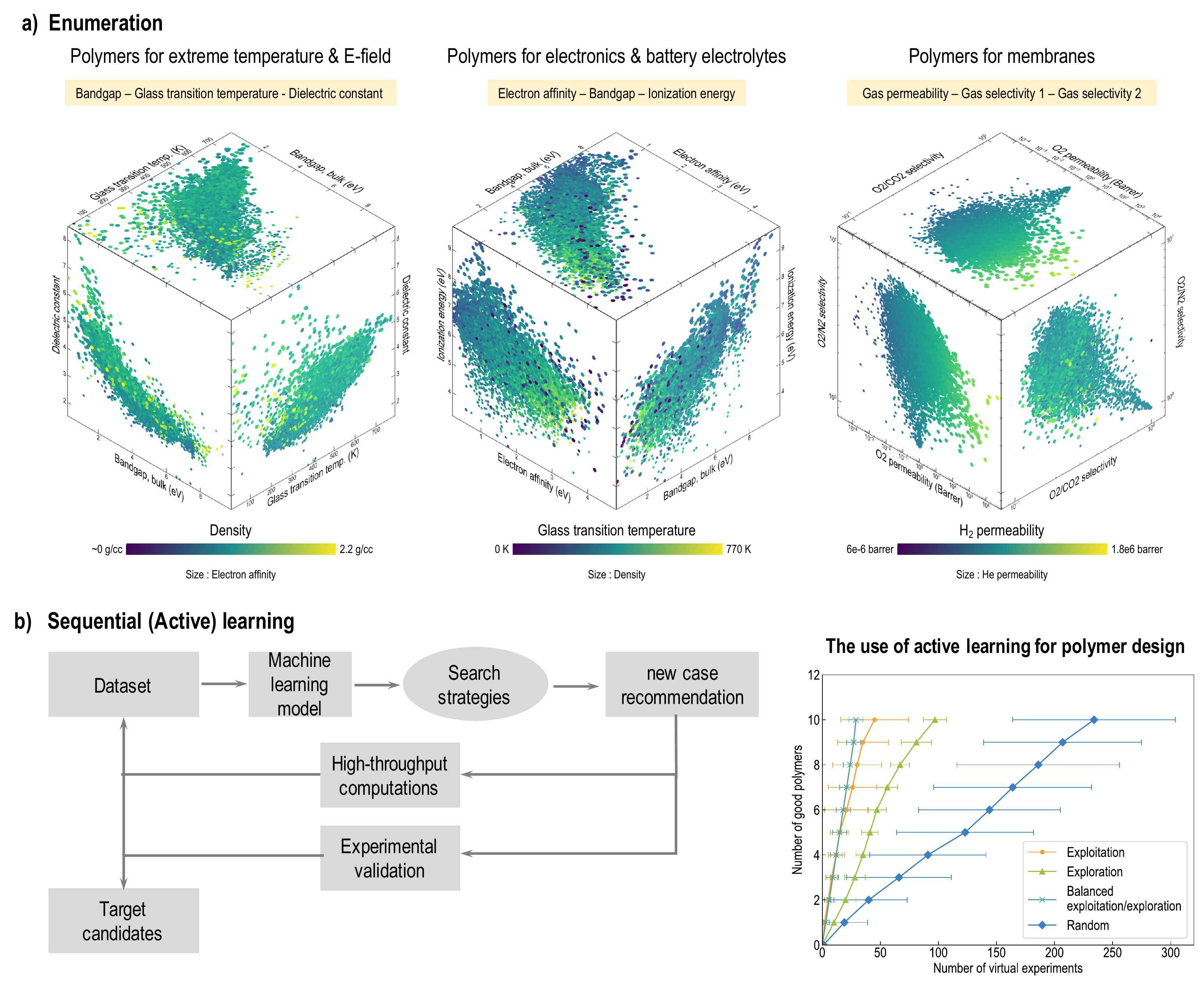}
		\caption{a) ML predicted properties of ten-of-thousands of enumerated known/synthesizable polymers. Different combinations of properties can be selected to screen polymers for the specific application, for example, large band gap, high glass transition temperature and dielectric constant for polymers tolerant to extreme temperature and electric field. b) Sequential (active) learning workflow (left) and its use for polymer design (right): number of experiments required (on average) to discover 1 –- 10 polymers with glass transition temperature greater than 450 K when starting with an initial dataset size of five polymers. The average is calculated using 50 different runs and the standard deviation is denoted by the error bar. This figure is taken from Ref. \cite{Kim2019-active} with permission from Cambridge University Press.}
		\label{en-ac}
	\end{figure*}

One of the dominant applications of machine learning techniques is to significantly accelerate the rational design and discovery of new materials by efficiently searching a pre-determined chemical space. The previously discussed ML models are used to predict the properties of a large pool of candidate polymers enumerated based on some physically or chemically motivated scheme, followed by a down-selection procedure based on certain screening criteria. The end result is a rank-ordered list of promising candidates for the target application.  The initially enumerated candidates may be previously synthesized polymers, or hypothetical polymers made by human experts or machine (e.g., genetic algorithm). Figure  \ref{en-ac}a) shows the trends, in a form similar to ``Ashby plots", of various ML predicted properties (such as glass transition temperature, band gap, dielectric constant (at THz) and density) for ten-of-thousands of known/synthesizable polymers.  These synthesized polymers have been manually accumulated from various resources, as discussed in Section \ref{data}. Depending on the property requirements for specific applications, different combinations of properties can be selected. For instance, polymers that are tolerant to extreme temperatures require large band gap, high glass transition temperature and dielectric constant, whereas polymers electrolytes used in Li-ion batteries require desired electron affinity, band gap, and ionization energy. Polymer membranes, on the other hand, require suitable gas permeability-selectivity pairs.

Following this forward design pipeline, Mannodi Kanakkithodi and co-workers identified promising polymer dielectrics with desired dielectric constant and band gap from a series of human-designed hypothetical polymers, made up of 4, 6 and 8 building blocks (e.g., --CH$_2$--, --CO--) \cite{Arun:design}. Likewise, Chen \textit{et al.} proposed five representative polymer candidates satisfying high glass transition temperature and required dielectric constant for high temperature, energy density capacitors and microelectronic devices from a pool of synthesized polymers \cite{chen2020frequency}. Additionally, Wu \textit{et al.}, on the other hand, used a surrogate thermal conductivity model based on transfer learning to screen promising candidates with target glass transition and melting temperatures, resulting in the synthesis of polymers with thermal conductivities of 0.18 -- 0.41 W/mK \cite{Wu-thermalconductivity}. Another successful example is from Kumar \cite{kumar}, wherein two polymer membranes with excellent CO$_2$/CH$_4$ separation performance were discovered from over 11,000 homopolymers, guided by GPR based gas permeability prediction models.  These findings strongly advocate the success of machine learning assisted forward design approach to discover polymer candidates for specific applications. 

\subsection{Sequential (Active) learning}

The polymer design algorithms discussed above provide a subset of promising polymer candidates with tailored properties for further validation via experimental synthesis or high-fidelity computations. However, these models are passive, with inherent errors in the property predictions owing to the limitations, such as bias or limited size of the training data. Thus, how one can dynamically and efficiently optimize polymers for the next experiment (or computation) is an important problem in materials discovery. It is far from trivial to select optimal candidates based purely on human intuition. In recent years, active-learning algorithms that exploit Bayesian optimization frameworks have been developed to effectively guide experiments or high-throughput computations for materials design, e.g., optimizing GaN LED structures, BaTiO$_{3}$ based piezoelectrics, and other inorganic materials for thermoelectric and electronic devices \cite{lookman2019active}.

As illustrated in Figure \ref{en-ac}b), active learning algorithms consist of three important components: 1) a surrogate model for the target property prediction; 2) an acquisition function to select the optimal point for the next experiment; 3) addition of the newly performed experiment to the knowledge dataset \cite{lookman2019active,Kim2019-active}.  The surrogate models in part 1) can be trained using various algorithms introduced in Section \ref{predictionmodel}.  To provide both prediction and uncertainty values of the target property, Gaussian process-based algorithms are common approaches used in active learning. There are other methods, such as support vector regression or decision trees in combination with bootstrapping methods, that estimate both the target property and its uncertainty. In part 2), the user can search unlabeled data by either  using the prediction uncertainties (called exploration), or by maximizing the target prediction values (called exploitation), or by balancing between exploration and exploitation. In the last part, the newly generated data from the new experiment is supplemented into the knowledge dataset to retrain the surrogate model. The whole pipeline is repeated until the target candidate is achieved.

In the polymer domain, Kim \textit{et al.} benchmarked the use of active learning to efficiently search polymers with $T_{\rm g}$ $>$ 450 K \cite{Kim2019-active}.  Figure \ref{en-ac}b) illustrates the average number of experiments required to discover 1 –- 10 polymers with a glass transition temperature of above 450 K, starting from an initial training dataset of 5 polymers. The error bars denote the standard deviation across the 50 different runs.  We note that on average 30, 46, 98 and 234 experiments were required to discover 10 high-glass transition temperature polymers using acquisition function definitions based on balanced exploitation and exploration, exploitation, exploration and random approaches, respectively. These findings indicate that the balanced exploitation and exploration method showed the best performance in terms of discovering promising polymer candidates.  Additionally, Huan \textit{et al.} have applied the active learning to automatically select polymer candidates for high-throughput DFT computations and find polymer dielectrics with a large band gap. More details are shown in Section \ref{polymerdesign}.   It is evident that the integration of active-learning within the materials discovery pipeline can guide materials design and dataset expansion in an efficient and targeted fashion.

\begin{figure*}[h]
	\begin{center}
		\includegraphics[width=1.0\textwidth]{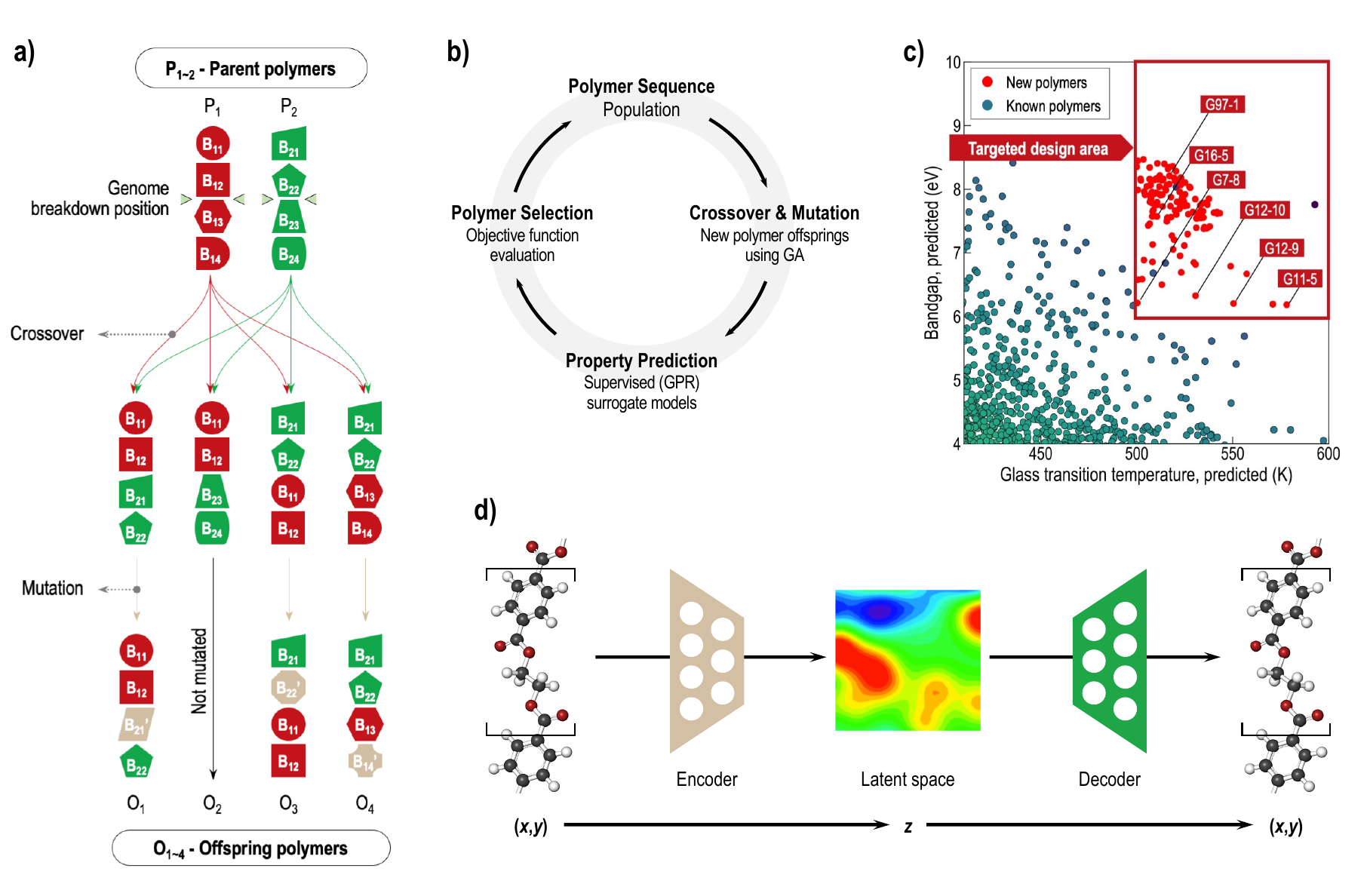}
		\caption{Polymer inverse design using machine learning. Use of evolutionary algorithms, such as GA, for design of polymers with target properties; a) basic operations of crossover and mutation to generate polymer offsprings, taken from Ref. \cite{kim2020polymer} with permission from ELSEVIER Publications;  b) the different stages of the iterative evolutionary process, involving population of new candidates, evaluation of fitness function using property-prediction models, and selection of the next generation with best fitness, taken from Ref. \cite{pilania_machine-learning-based_2019} with permission from ACS Publications; c) results for an exemplary polymer design problem of high glass transition temperature and large band gap, taken from Ref. \cite{kim2020polymer} with permission from ELSEVIER Publications; d) Use of variational autoencoders (VAE) for polymer design. The latent space is searched to find polymers with desired properties, which are ‘generated’ using the decoder mapping.    }
		\label{VAE-GA}
	\end{center}
\end{figure*}	

\subsection{Evolutionary strategies}
Another approach to polymer discovery is the ``inverting the prediction pipeline".
Contrary to the enumeration approach that relies on virtual screening of polymers from a pre-defined candidate set using instantaneous property prediction from surrogate models, inversion problems focus on directly generating polymers that satisfy given property objectives, making it a more general approach to materials discovery. Two approaches for direct materials design have emerged: first, the use of evolutionary methods, such as the genetic algorithm (GA) \cite{Huan:design,kim2020polymer}, and second, generative ML approaches, such as variational autoencoders (VAE) \cite{kingma2014stochastic} and generative adversarial networks (GAN) \cite{goodfellow2014generative}. We describe the evolutionary approaches here, while generative methods are discussed in Section \ref{generative}.

GA is based on the principle of natural selection. The inherent structure of a polymer makes its treatment using GA straightforward—a polymer can be thought of as a sequence of chemical building blocks (e.g. CH$_2$, C$_6$H$_6$, or blocks B$_{12}$, B$_{13}$ in Figure \ref{VAE-GA}a) connected to each other by covalent bonds (analogous to DNA base pairs), and the properties of a polymer are functions of the sequence of constituent chemical building blocks (analogous to how oculocutaneous albinism II (OCA2) gene sequence mostly dictates human eye color). Within the GA approach a series of \emph{crossover}, \emph{mutation} and \emph{selection} operations are applied to discover new candidate polymers with desired properties. It starts with a random generation of (say, 100) polymers, whose chemical building blocks are modified using crossover—pruning and mixing of monomer building blocks—and mutation—random alterations to monomer building blocks—operations to obtain a large set of offspring polymers, as illustrated in Figure  \ref{VAE-GA}a). Next, the top offspring candidates with desired properties are selected based on their user-defined objective score to form the next generation of polymers. This GA cycle is repeated until a sufficient number of candidates with desired properties are obtained, as in Figure \ref{VAE-GA}b). Besides polymer discovery, GA has also been utilized to solve other problems in materials science, such as developing functional forms of interatomic potentials \cite{hernandez2019fast}, and discovering hidden material property relations \cite{gandomi2016genetic}.

A critical component of the GA design scheme is the evaluation of the objective function during the selection stage. This step has been a major bottleneck for polymer discovery since property estimation through experiments or computations is very expensive and time-consuming \cite{fjell2011optimization}. However, with the recent development of cheap and reliable polymer property models (Section 5.1), the objective function can now be computed in a fraction of a second. This allows the GA process to truly explore a very rich chemical polymer space, going well-beyond the pre-defined candidate sets. Furthermore, by setting-up a property weighted objective function, polymers that simultaneously satisfy multiple property criteria can be targeted.
	
Kim and co-workers used GA to design polymers with high glass transition temperature and large band gap, which are useful for high-energy capacitors because of their stability at both high temperatures and electric fields \cite{kim2020polymer}. Notably, this is a difficult design problem, with only 4 out of thousands of known polymers displaying glass transition temperature $>$ 500 K and band gap $>$ 6 eV. Two interesting aspects of their design process was the choice of the chemical building blocks, and the underlying property prediction models. The former consisted of a comprehensive list of 3,045 chemical blocks, extracted from $\sim$ 12,000 synthetically known polymers using the concept of BRICS---similar to the hypothetical polymer design work in Section \ref{data}. Each block had 1 -- 4 connection points that were used to form/break bonds with other chemical blocks during the crossover and mutation operations. For the latter glass transition temperature and band gap prediction models, they used two independent GPR surrogate models based on a hierarchical polymer fingerprinting scheme (Section \ref{feature}) that were trained on an experimental and DFT computed dataset of 5,072 and 562 polymers, respectively. During 100 generations of the GA cycle, they successfully designed 132 new polymers that meet the target properties, as opposed to only 4 previously known cases (see Figure  \ref{VAE-GA}c)). Furthermore, their analysis of the identified polymer candidates revealed insights about the key fragments that promote high glass transition temperature and large band gap in polymers, such as the presence of terminal difluorocarbon or trifluoromethyl, saturated 5-or 6-membered rings, oxolane, etc. These findings are compatible with known chemistry. For instance, fluorine atoms induce large band gap through the formation of lower (higher) C--F sigma bonding (anti-bonding) orbitals. A similar approach has been utilized to design polymers with high dielectric constant, although it considered a relatively small number of possible chemical building blocks \cite{Arun:design}.
	
In a different study, Pilania \textit{et al.} used GA to design bio-advantaged (biosynthesizable and biodegradable) Polyhydroxyalkanoate (PHA)-based polymers with desired glass transition temperature values \cite{pilania_machine-learning-based_2019}. A machine learning model trained on an experimentally-measured and carefully-curated glass transition temperature values for a wide range of PHA homo- and co-polymers were combined with a GA-based search and optimization routine to explore a much wider chemical space formed by multi-component polymer chemistries, beyond co-polymers. Furthermore, by explicitly integrating the prediction uncertainties and number of polymer components in the GA objective function, they were able to focus their search on polymers containing desired number of components (ternary, quaternary, etc.) where the confidence level in the machine-learned glass transition temperature predictions were higher than a pre-specified cutoff value.

\subsection{Generative models}\label{generative}

Based on the concept of unsupervised learning, VAE and GAN offer a different route for targeting inverse polymer design. They learn a mapping from a continuous latent space to the polymer space, using which new candidates with desired properties are generated after solving the optimization problem in the more amenable latent space. For example, in the case of VAE, the encoder unit learns to represent a polymer in a high-dimensional (say, 100 -- 200) continuous (latent) space, while the decoder unit learns to map back a vector in the latent space to a valid polymer, as shown in Figure \ref{VAE-GA}d). Both mappings are important from a materials design perspective: the encoder provides a fingerprinting scheme that can be exploited by different “forward models”, while the decoder provides a pathway to systematically search polymers in a proxy latent space using different optimization schemes, and later generate the desired polymer  candidate associated with the optimal latent point. Although the VAE and GAN approaches have received attention for molecule or drug discovery \cite{VAE1,VAE2,VAE3,GAN2,GAN3}, they are only beginning to be exploited for extended systems such as polymers.
	
A key challenge in developing such a generative model is that the decoder unit should map a continuous latent space to a discrete and structured material space, which should represent a valid candidate material as dictated by chemistry. For example, in case of polymers, the decoder output should necessarily have two chain ends, or the involved C and O atoms should display a valency of 4 and 2, respectively. This goal of enforcing the decoder to output valid polymers was recently achieved by Batra $et~al.$ \cite{batra2020polymer} using syntax-directed VAE. The strategy involves representing the polymers using their SMILES representation, and then imposing the decoder to obey both the syntactic and semantic constraints associated with the class of polymers; syntactic refers to the grammatical rules inherent to the SMILES language, while semantic refers to the contextual constraints driven by polymer chemistry. The inclusion of explicit syntax and semantics in the VAE model improves the quality of the learned latent space. It also leads to a high occurrence of valid polymer SMILES upon decoding, making the process of discovery efficient.

Batra $et~al.$ coupled the unsupervised syntax-directed VAE with the supervised GPR method to discover polymers with high glass transition temperature and large band gap \cite{batra2020polymer}. They used the encoder unit to fingerprint the polymers, which were then mapped to the respective glass transition temperature and band gap values using GPR. To train the VAE model they had to overcome a crucial data sparsity challenge: to-date the total number of chemically diverse polymers synthesized is $\sim$ 12,000, while a VAE model usually requires $>$ 100,000 points for its training. They used retro-synthetic ideas to generate a representative hypothetical dataset of $\sim$ 250,000 polymers, constructed from the previously discussed set of 3,045 chemical building blocks. For the discovery of polymers with target properties, they first encode a few known polymers that satisfy the given design criteria to find regions in the latent space where desirable polymers are expected to be present. Linear interpolations within these preferred regions of the latent space are then used to select latent points for which GPR property predictions meet the desired goals. Finally, the decoder is used to obtain the polymer SMILES associated with such selected latent vectors. Several hundreds of new polymers that meet the target property objectives were generated using this process. We anticipate that the concepts of transfer learning, multi-task learning and semi-supervised learning will advance the use of generative models for polymer discovery.

The following comparisons between the GA and the generative techniques for inverse design can be made. First, GA is relatively easy to interpret and entails little efforts to tune the involved parameters (mutation chance, initial population, etc.). In contrast, VAE models being based on NNs are almost impossible to interpret and often entail hefty parameter tuning efforts. Second, the space explored by GA is somewhat constrained by the polymer building blocks, but the SMILES based polymer representation allows VAE to explore a much wider chemical space in a truly unconstrained manner. Lastly, prior knowledge can be easily incorporated in GA, for instance, by biasing the initial population and/or the mutation operation towards favorable building blocks. However, more comparative studies would be needed in the future to establish methods that are appropriate under different scenarios.

	\section{Application examples}\label{applications}
 Polymers are useful in a range of applications. To be a good candidate for any  specific application, they must meet multiple desired property requirements, as summarized in Table \ref{tab:application}  for selected applications. Below, we comment on a few such applications, with an emphasis on key properties relevant for those applications which may be used to formulate screening criteria (also captured in Table \ref{tab:application}).
 
\begin{table}[]
    \centering
\caption{Desired properties of polymer candidates for various applications}
    \begin{tabular}{lll}
    \hline
    \hline
       Applications   &Representative desired polymer properties\\
       \hline
       Capacitors& Large band gap and high charge injection barriers\\
       (polymer dielectrics)&  high glass transition temperature and dielectric constant\\
       \hline
      Li-ion batteries& Large electrochemical stability window, high ionic conductivity, \\
      (polymer electrolytes)& high Li-ion transference and mechanical strength\\
      \hline
      Polymer membrane& High permeability and selectivity for gas pairs\\
      \hline
      Electronic devices
      (conducting polymers) & High electrical conductivity\\
    \hline
    \hline
    \end{tabular}

    \label{tab:application}
\end{table}

	\subsection{Polymer dielectrics design for high energy density capacitors}\label{capacitor-part}

Polymer-based dielectric capacitors are widely used in energy storage devices \cite{Chu334,liqi,qitan,sharma2014rational,HUAN2016236,ho2018polymer, PONB}. Given the increasing needs of high energy density capacitors, the development of polymer informatics can significantly facilitate the discovery of novel polymer dielectrics \cite{sharma2014rational,mannodi2016machine, mannodi2016rational,Huan:Data,MANNODIKANAKKITHODI2017}. Typically, good polymer dielectrics for high energy density capacitors need to satisfy several property requirements, e.g., high dielectric constant and high breakdown strength (which is positively correlated with band gap and charge injection barriers of metal/polymer interfaces \cite{deepak-interface}).  Further, polymers with high glass transition temperature are desired for high-temperature capacitors to enhance the thermal stability at extreme temperature \cite{PONB,ho2018polymer}. Thus, the criteria of high glass transition temperature and $\epsilon$, large band gap and high charge injection barriers can be utilized, in combination with machine learning, to screen polymer candidates tailored to extreme high-temperature and electric field.  For instance, several representative dielectrics films with high dielectric constant and band gap, have been successfully designed and synthesized using computation- and data-driven strategies \cite{mannodi2016rational,MANNODIKANAKKITHODI2017}. Additionally,  many representative polymer dielectrics are being proposed for high-temperature capacitors by either screening known/hypothetical polymers using the enumeration method \cite{deepak-interface,chen2020frequency} or using the generative models, such as GA \cite{kim2020polymer} and VAE \cite{batra2020polymer}, as described in Section \ref{polymerdesign}.
	 
 	\subsection{Polymer membrane design for gas separation}
Polymers are also promising candidates for gas separation due to their high surface area \cite{low2018gas,jue2015targeted,kumar}. A typical class of polymers called polymers of intrinsic microporosity, has attracted great attention since the early 1990s \cite{low2018gas}. The present polymer membranes suffer from low selectivity and physical aging, calling for the exploration of novel polymeric membranes. However, it is non-trivial to find promising polymer membranes with a combination of high permeability and selectivity (or above the upper bound of ``Robeson plots" \cite{robeson2008upper}) for different gas pairs, e.g., O$_{2}$/N$_{2}$. Some initial attempts have been performed to  speed up the polymer membrane search using data-driven approaches, for instance, building gas permeability prediction models \cite {zhu2020polymer} (see Section \ref{predictionmodel}) and identifying polymer membrane candidates for CO${_2}$/CH${_4}$ separation using the enumeration method \cite{kumar}. 
 
\subsection{Polymer electrolytes design for Li-ion batteries} 
Rechargeable Li-ion batteries have been widely adopted in many applications from micro-electronics to aerospace. Motivated by their commercial need, the development of novel and safer solid polymer electrolyte materials has caught ever-increasing attention \cite{sequeira2010polymer,agrawal2008solid,mindemark2015high,electrolytesreview}. To optimize the performance of Li-ion batteries, the polymer electrolytes should have a wide electrochemical stability window, high ionic conductivity and Li-ion transference, and low glass transition temperature. Since it is time-consuming to search optimal electrolytes using experiments,  data-driven aided polymer design strategies provide a great opportunity. For example, we previously noted that Wang \textit{et al.} designed novel polymer electrolytes with high ionic conductivity using machine learning aided coarse-grained molecular dynamics simulations \cite{SPE-ML}. Additionally, the property prediction models (Section \ref{predictionmodel}) and the design algorithms (Section \ref{polymerdesign}) discussed above are powerful methods to screen/design polymer electrolytes satisfying multiple property requirements.

\subsection{Conducting polymers design for electronic applications} 
Although polymers are usually insulators,  there is a class of intrinsically conducting polymers used in electronic devices, such as light-emitting diodes, field‐effect transistors and organic solar cells \cite{mayer2007polymer,hofmann2018highly,brebels2017high}.  Molecular doping is often used to further increase the conductivity of polymers \cite{hofmann2018highly}, but it slows down the discovery of optimal polymer-dopant pairs with high conductivity because of the complex nature of the electron transfer mechanisms, dopants and polymers interactions, and processing conditions.  This situation can be improved using polymer informatics, e.g. developing conductivity prediction models and screening optimal polymer-dopants pairs using the enumeration method.  It is supported by the discovery of several high-performing donor/acceptor pairs for organic solar cells using random forest and boosted regression trees based property prediction models \cite{wu2020machine}.

\subsection{Biodegradable and depolymerizable polymers discovery}

Bioplastics, such as those derived from plants and bacteria, are rich in highly oxygenated molecules. They can be utilized in the production of monomers capable of facile conversion to plastic materials that are easily degradable in the environment \cite{albertsson_designed_2017}. However, to fully harness the power of these nonconventional biosynthesis routes, it needs to establish structure-property relationships to identify desired application-specific optimal chemistries. To understand this problem better, Pilania \textit{et al.} have proposed a machine learning route to learn structure-property mappings in PHA-based polymers 
from polymer data \cite{pilania_machine-learning-based_2019}. Moreover, it is critical to discover new biodegradable polymer candidates with high biodegradability. Because low crystallinity, melting temperature and glass transition temperature lead to large amorphous regions and favor biodegradation, ration-design of biodegradable polymers satisfying these properties using data-driven methods can be an important research topic. For instance, some new biodegradable polymers with desired glass transition temperature have been designed recently  using GA \cite{pilania_machine-learning-based_2019} (see Section \ref{polymerdesign}).

Additionally, depolymerizable polymers are playing an increasingly important role in practical applications, especially in drug delivery, recyclable plastics, self-healing and recyclable coating materials \cite{kaitz2015depolymerizable,dilauro2013reproducible}. Such great interest is motivated by the fact that depolymerizable polymers, upon exposure to particular stimuli, can be triggered to rapidly depolymerize into monomers at moderate to relatively low temperatures. As a result, polymers with low ceiling temperatures are desirable, where ceiling temperature is the temperature at which the polymerization and depolymerization rates are in equilibrium. Because of the limited available number of known polymers with low ceiling temperatures, it is greatly desired to propose computational strategies to estimate the ceiling temperature. Further, the data-driven design tools involved in polymer informatics could be applied to rapidly screen such depolymerizable polymers.

\section{Critical next steps}\label{nextstep}
\subsection{Beyond homopolymers}
So far, many data-driven approaches have been limited to homopolymers. The space of co-polymers, polymer blends and polymers with additives/nanocomposites remains largely unexplored but has great practical significance. Brinson and co-workers have spent significant efforts to develop ``NanoMine" for polymer nanocomposites analysis and design \cite{zhao2016perspective}. However, it is still non-trivial to treat these types of polymers, because of their complicated chemical and physical structures. Co-polymers consist of two/more monomer or basic building unit  types, and could be branched or linear co-polymers (further classified as block, alternating and random co-polymers based on the structural arrangement of different monomers). Polymer blends are mixtures of two or more polymers, including homogeneous, immiscible and heterogeneous polymer blends.  The ratios and structural arrangements of different monomers (or polymers) significantly impact properties of co-polymers and polymer blends, but only sparse data is available on this topic. Moreover, it is challenging to systematically and dynamically collect such data from various resources.  Thus, advanced techniques need to be developed to collect, represent and learn data of more complex varieties of polymers.

\subsection{Sustainable data capture}

The core requirement for polymer informatics is a broad-based data acquisition and management infrastructure. In addition to the limited number of available polymer databases and polymer handbooks, a large amount of scientific data remains untapped in numerous scientific journals, including text, tables or figures. While the manual text excerption of such journals is very time consuming and laborious, machine learning-based NLP methods are more powerful and promising tools to expedite and automate this process. The application of NLP tools in material science is still in its infancy. More efforts are needed to incorporate materials or polymers domain knowledge into existing NLP algorithms (e.g., word2vec \cite{mikolov2013efficient}) to train word-vectors (numerical vectors that represent distinct words) for scientific information retrieval. To achieve this goal, Named Entities Recognition (NER) is the most important step, i.e., tokenizing the words into scientific meanings (e.g., chemical species, synthesis conditions and characterization methods). ChemDataExtractor \cite{ChemData}, ChemSpot \cite{Chemspot} and ChemTagger \cite{ChemTagger} are available toolkits for extracting chemical information of materials from scientific articles, such as inorganics and molecules. Similar tools need to be developed for the polymer domain.

However, polymers pose additional challenges \cite{adams2008engineering,soft-informatics}, as there is no standard or complete polymer name entity dictionaries. A collection of source-based, structure-based, traditional and abbreviation names are interchangeably used to name polymers \cite{adams2008engineering}.  For example, polyethylene is also called PE, poly(ethylene) and poly-(ethylene), but all these possible names should be treated as the same entity (in a process referred to as ``normalization" by the NLP community). In addition to names, more efforts are required to assign polymer notations for specific categories, e.g., properties, synthesis recipes and characterization technologies. Therefore, it is of great importance to create unique and standard polymer related dictionaries in the future.   Other important issues include building efficient toolkits to interpret monomer SMILES from polymer names, identifying structure (or polymer names)-property relationships in texts, and extracting valuable material property contained in images and tables. 

\subsection{Polymer representation and learning}
As discussed in Section \ref{feature}, molecular or polymer-based fingerprints can provide acceptable prediction accuracy for many polymer properties, e.g., glass transition temperature, band gap, dielectric constant and gas permeability. This is because the chemical structure of the monomers plays a dominant role in determining these properties. However, other important polymer properties, including crystallinity, mechanical properties (e.g., tensile strength) and solution behavior, strongly depend on their molecular weights, morphologies (linear or cross-link) and processing conditions (temperature, pressure and cooling-rates). Incorporating these descriptors in the fingerprint framework is critical to the creation of accurate,  robust and universal property prediction models.

In addition to enriching the polymer fingerprint definition, more advanced neural networks algorithms can be applied for learning the latent knowledge, property prediction and polymer generations.  For instance, the transfer-learning or multi-task learning approaches have great potential to deal with the sparse data issue in polymers. The former modifies the latent features learned using one source task to learn a different  target task, while the latter trains multiple source tasks and the shared features used to learn a target task. The key concept common between these two methods is the learning of a shared (polymer) representation between related properties (or materials). These algorithms have been successfully applied in the domain of drug design and bioinformatics \cite{Ramsundar2015,Ramsundar2017,Ma2015}. In polymers, as mentioned in Section \ref{predictionmodel}, Yamada \textit{et al.} have used the transfer-learning method to predict properties of polymers using pre-trained models of molecules and inorganic materials \cite{transfer}.  However, large and diverse datasets of related property (tasks) are essential for the success of such models, as only then there is a high chance of learning transferable features and achieving accurate predictions for the target task.

Another important topic is the use of graph neural networks (GNN) in polymer informatics. In contrast to traditional manually designed fragment-based ML models, GNN represents materials as graphs (typically, atoms as nodes and their bonds as edges) and automatically find their optimal fingerprint representation depending on the downstream learning task, leading to its wide applications for molecules.  However, the use of GNN for polymers has been limited \cite{zeng2018graph,sumpter1994neural} owing to the difficulty in treating large-scale polymers using graphs. Further, polymers are made up of numerous repeat units, and the best way to treat connection points between repeat units in a graph is unclear. Using oligomers to replace polymers is a potential solution \cite{sumpter1994neural,zeng2018graph}, however, its prediction capability needs to be tested. Additionally, ideas on graph generative methods for molecules, e.g., atoms- and substructure-based encoder-decoder methods \cite{jin2020hierarchical} could be extended for polymers using GNN. Another interesting approach of graph-to-graph translation was recently put forth to optimize molecules with desired properties, by assembling one molecular graph with another of the target properties \cite{jin2018learning}. All of these techniques can be adapted for polymers, provided the following challenges are addressed. Many polymers have large-sized monomers ($>$ 50 atoms), making it difficult to correctly assemble potential fragments during the decoding process.   Motifs-based methods can greatly increase the reconstruction and validation accuracy for polymer generation by using large-size motifs as building blocks \cite{jin2020hierarchical}. However, other concerns remain, such as chemical or thermodynamic stability of generated polymers and their synthetic feasibility.

 \subsection{Polymer retro-synthesis planning}
Even with the knowledge of which polymer to make for a given application (designed, for instance, using intuition, computation or machine learning), the realization of the polymer can still be very slow because of  synthesis challenges. There are various uncertain factors, such as unavailability, toxicity  or high cost of the raw materials or demanding technical steps. In the past, the synthesis pathways adopted for a target polymer have been heavily dependent on the domain knowledge and personal preferences of experimenters. Computer-assisted retro-synthesis techniques have been widely developed in the last several decades to identify a series of reaction pathways leading to the synthesis of a target product. In the domain of molecules, either template-based \cite{segler2018planning,coley2017prediction} or template-free \cite{jin2017predicting,C8SC04228D,NIPS2019_9090} machine learning approaches have been built for product prediction and have achieved promising results. However, no such method has been developed yet for the case of polymers. Complications in the polymer synthesis processes, e.g., various polymerization mechanisms (such as addition, ring-opening and condensation polymerization), the selection of optimal monomers and solvent pairs, processing conditions (such as cooling rates or annealing temperatures) will need to be considered. Further, unlike molecules, there is no library of reaction templates for polymerization. Nevertheless, experimental polymer synthesis data is plentiful, which can be accumulated manually or using NLP methods, and processed appropriately to develop machine learning models for polymer synthesis and retro-synthesis planning.

\subsection{Autonomous integration of experimental and computational workflows}
As all the different pieces of AI-assisted chemical search, retro-synthesis planning and processing optimization come together, the idea of autonomous polymer synthesis and design is expected to become a reality. In fact, examples of autonomous robot researchers with the ability to synthesize drugs for tropical diseases \cite{williams2015cheaper}, carbon nanotubes with targeted growth rates \cite{nikolaev2016autonomy}, layered superlattices \cite{masubuchi2018autonomous}, and even perform x-ray scattering measurements \cite{noack2019kriging}, have already been demonstrated recently.
However, polymers owing to their structural, chemical and processing complexity pose unique challenges for autonomous design.  For instance, the average molecular weight of a polymer, which predominantly dictates its properties, is highly sensitive to the processing time and conditions. Learning such complex relations, from the data alone, will be challenging for an autonomous researcher. The real-time/in-line characterization of polymers is also difficult owing to their complex semi-crystalline/amorphous structure, or due to the different degree of branching or stereochemical relationships. Nonetheless, the prowess of autonomous labs in terms of time and cost benefits, experimentation consistency, long hours of operation, and efficient and robust search of parameter spaces is expected to guide polymer discovery in the future.

\section{Acknowledgments}
This work has benefited from the generous support by the Office of Naval Research, the Toyota Research Institute, the Department of Energy and the National
Science Foundation on machine learning related research through several grants. G. P. acknowledges support from the Laboratory Directed Research and Development (LDRD) program of Los Alamos National Laboratory
under the BioManIAC project \# 20190001DR. C. Ku. acknowledges support from the Alexander von Humboldt Foundation. R.B acknowledges support by LDRD funding from Argonne National Laboratory, provided by the Director, Office of Science, of the U.S. Department of Energy under Contract No. DE-AC02-06CH11357, and the use of the Center for Nanoscale Materials, an Office of Science user facility, supported by the U.S. Department of Energy, Office of Science, Office of Basic Energy Sciences, under Contract No. DE-AC02-06CH11357. Joseph Kern is gratefully acknowledged for a critical reading of the manuscript.

%\bibliography{ref1}

\end{document}